\documentclass[aps,prl,twocolumn,letterpaper,superscriptaddress,showpacs]{revtex4}
\usepackage{graphicx}
\usepackage{CJK}

\begin{document}
\begin{CJK*}{UTF8}{bsmi}
\title{
Do Transition Metal Substitutions Dope Carriers in Iron Based Superconductors?
}
\author{Tom Berlijn}
\affiliation{Condensed Matter Physics and Materials Science Department,
Brookhaven National Laboratory, Upton, New York 11973, USA}
\author{Chia-Hui Lin (\CJKfamily{bsmi}林佳輝)
}
\affiliation{Condensed Matter Physics and Materials Science Department,
Brookhaven National Laboratory, Upton, New York 11973, USA}
\affiliation{Department of Physics and Astronomy, Stony Brook University, 
Stony Brook, New York 11790, USA}
\author{William Garber}
\affiliation{Condensed Matter Physics and Materials Science Department,
Brookhaven National Laboratory, Upton, New York 11973, USA}
\author{Wei Ku(%
顧威)
}
\affiliation{Condensed Matter Physics and Materials Science Department,
Brookhaven National Laboratory, Upton, New York 11973, USA}
\affiliation{Department of Physics and Astronomy, Stony Brook University, 
Stony Brook, New York 11790, USA}

\date{\today}

\begin{abstract}
We investigate the currently debated issue concerning whether transition metal substitutions dope carriers in iron based superconductors.
From first-principles calculations of the configuration-averaged spectral function of BaFe$_2$As$_2$ with disordered Co/Zn substitutions of Fe, important doping effects are found beyond merely changing the carrier density.
While the chemical potential shifts suggest doping of a large amount of carriers, a reduction of the coherent carrier density is found due to the loss of spectral weight.
Therefore, none of the change in the Fermi surface, density of states, or charge distribution can be solely used for counting doped coherent carriers, let alone presenting the full effects of the disordered substitutions.
Our study highlights the necessity of including disorder effects in the studies of doped materials in general.
\end{abstract}


\maketitle
\end{CJK*}

Doping is one of the most powerful tools for tuning the electronic properties of functional materials. 
Well known examples include n- and p-type semi-conductors, dilute magnetic semi-conductors and numerous heavy fermions and strongly correlated oxides (e.g. manganites, cobaltates, and cuprates.)
In most of these cases, inclusion of doped impurities are considered an effective way to introduce carriers, but recently the amount of carrier doping has become the center of a serious debate in the field of iron based superconductors (FeSCs).
A density functional theory (DFT) study~\cite{wadati} reported that the additional charge density in transition metal (TM) substituted FeSCs are fully localized at the TM dopant site, which lead~\cite{wadati} to the surprising conclusion that the TM substitutions do not dope carriers. 
While the lack of additional density at the Fe sites has recently been confirmed experimentally~\cite{bittar,merz}, this widely discussed conclusion of the system being undoped, however, seems to be in contradiction with several experimental~\cite{malaeb, neupane,liu} and theoretical~\cite{neupane,malaeb,nakamura,konbu} studies that indicate a change of Fermi surfaces (FSs) consistent with carrier doping.
A serious debate was thus initiated in the field: do the TM substitutions dope carriers into the FeSCs or not?

In this Letter we investigate this issue via first-principles DFT calculation of \emph{disordered} substitutions.
Contrary to the previous theoretical studies, which employed rather small periodic super cells, we compute the configuration-averaged spectral function of disordered TM substitutions in BaFe$_2$As$_2$, using the weak Co and strong Zn impurities as representative examples.
In the case of Co substitution, despite the localized charge distribution, no apparent localized states are found and a large chemical potential shift results from the injection of additional carriers to the system.
In the case of Zn substitution, while both strongly and weakly localized states are found, the chemical potential is still shifted noticeably, reflecting real carrier doping to the system.
However, an important loss of coherent spectral weight results from scattering against the disordered impurities.
Consequently, the actual number of coherent carriers in the system reduces noticeably from the Luttinger count of the FSs.
Our findings illustrate that none of the FSs, density of states, or the charge density can be used for a physically meaningful counting of coherent carriers in a disordered system in general.
More importantly, physical effects of the disordered impurities are much richer than simply adding carriers.
Specifically, the emergence of incoherent carriers and loss of coherent carriers holds interesting potentials to strengthen superconductivity over magnetism, and should be taken into account in future studies.

\begin{figure}
\includegraphics[width=1.0\columnwidth]{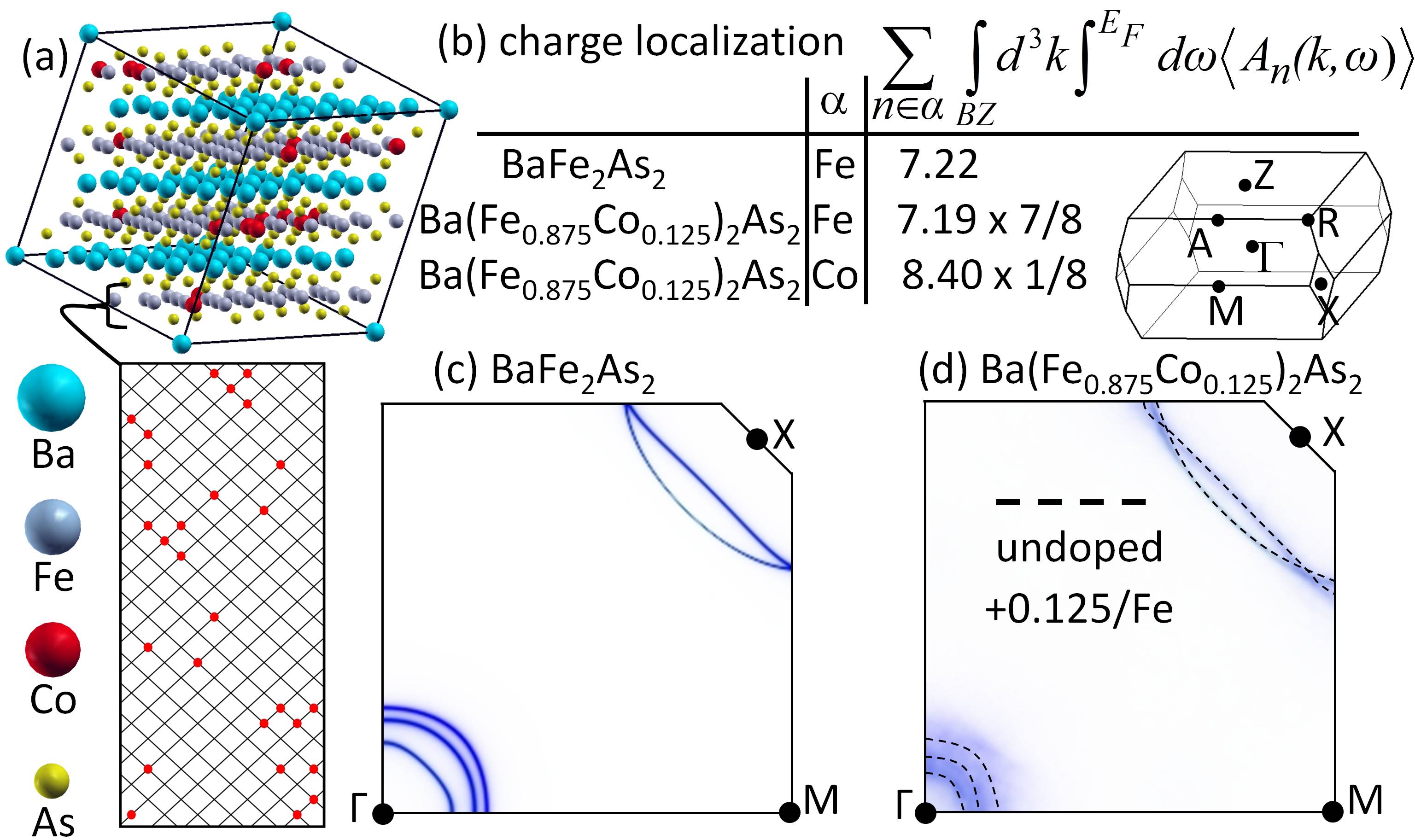}
\caption{\label{fig:fig1}
(color online) 
Two seemingly contradicting results: charge localization and carrier doping.
(a) An example of large sized supercells used for the configurational average.  
(b) The average occupation of the atomic Wannier orbitals showing the doped charge to be localized at the Co site. 
The FSs of (c) BaFe$_2$As$_2$ and 
(d) Ba(Fe$_{0.875}$Co$_{0.125}$)$_{2}$As$_{2}$ showing the hole pockets at $\Gamma$ to shrink and the electron pockets at X to grow, consistent with adding 0.125 electrons/Fe to the undoped system.
}\end{figure}

\begin{figure}
\includegraphics[width=1.0\columnwidth]{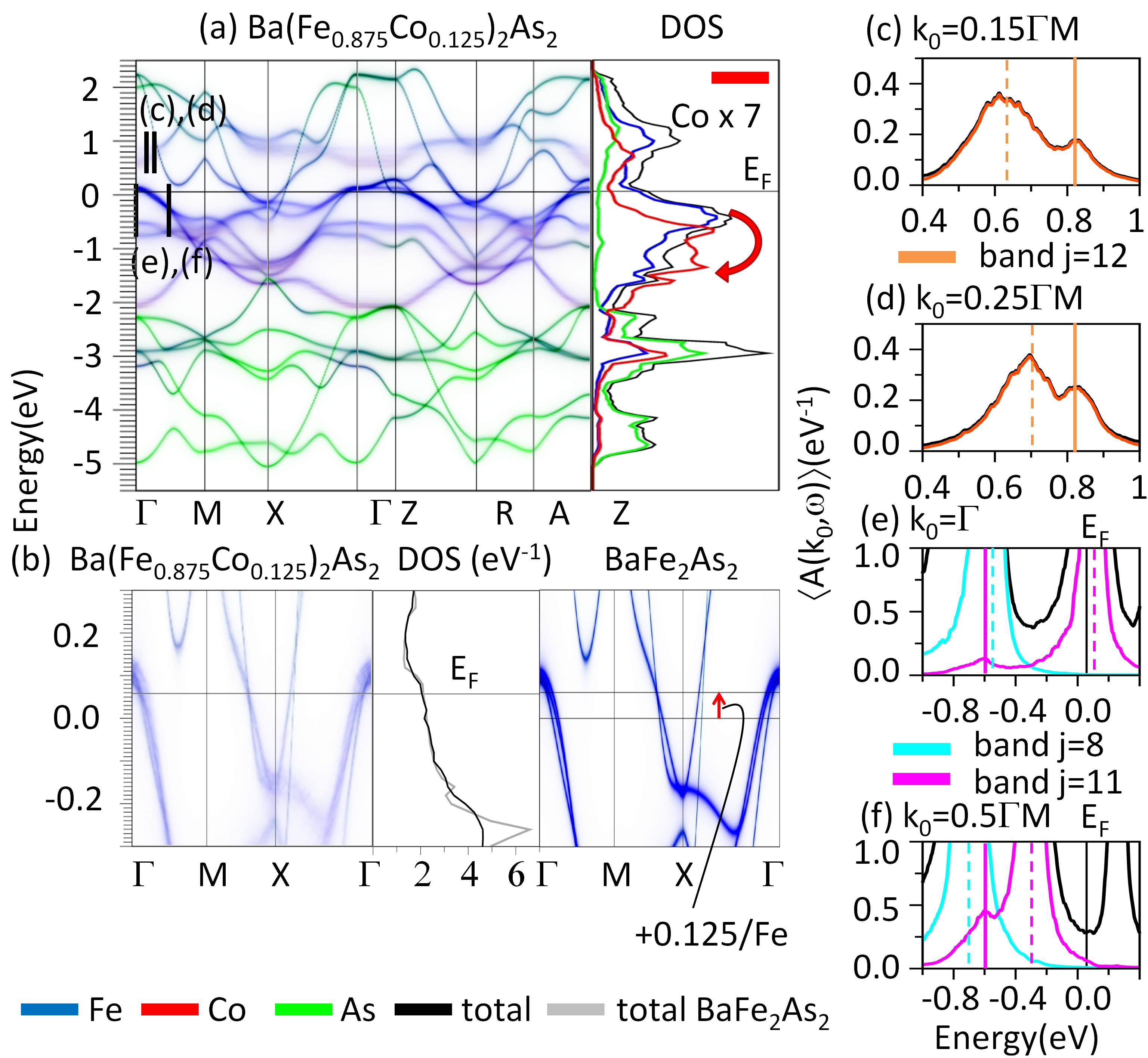}
\caption{\label{fig:fig2}
(color online)
Configuration averaged spectral function of disordered Ba(Fe$_{0.875}$Co$_{0.125}$)$_2$As$_{2}$ in the full low energy Hilbert space (a) where the partial DOS of Co is enhanced by a factor of 7.(b) Comparison of spectral function and DOS of disordered system (black) with the undoped one (grey) near the chemical potential. Bloch based spectral function of band $j=12$ at (c) $k_0=0.15\Gamma$M  and (d) $k_0=0.25\Gamma$M and of bands $j=8,11$ at (e) $k_0=\Gamma$ and (f) $k_0=0.5\Gamma$M, where the dashed/solid lines indicate the centers of the dispersive/non-dispersive components. The black curves in (c)(d)(e)(f) correspond to the total spectral function.}\end{figure}

\begin{figure}
\includegraphics[width=1.0\columnwidth]{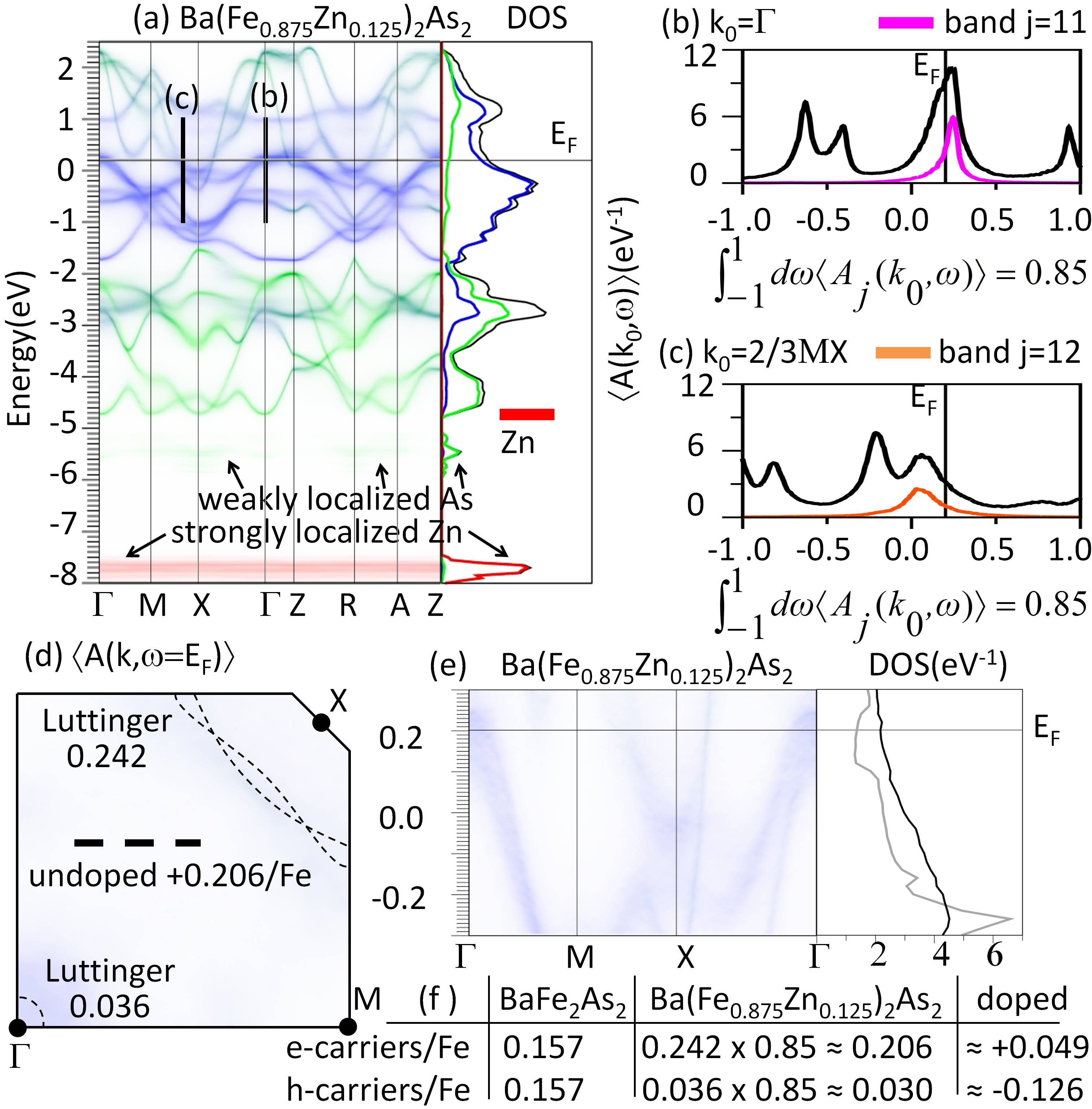}
\caption{\label{fig:fig3}
(color online)
Configuration averaged spectral function of disordered Ba(Fe$_{0.875}$Zn$_{0.125}$)$_2$As$_{2}$ in the low energy Hilbert space (a).
Bloch based spectral function (b) of band $j=11$ at $k_0=\Gamma$ and (c) of band $j=12$ at $k_0=2/3$MX.
(d) FSs of doped system compared with the undoped system with $\sim 0.2$ additional electrons per Fe. (e) Band structure and DOS of doped (black) and undoped (grey) system near the chemical potential. (f) The estimated number of doped coherent carriers. The black curves in (b)(c) correspond to the total spectral function.
}\end{figure}

The configuration-averaged Wannier based spectral function $\langle A_{n}(k,\omega)\rangle$ of Wannier orbital $n$, frequency $\omega$, and crystal momentum $k$ (in the 2-Fe Brillouin zone~\cite{lin}) is obtained by averaging over 10 large random-shaped supercells containing 400 atoms on average, including 12.5\% random TM substitutions (e.g. Fig.~\ref{fig:fig1}(a)). 
(For the spectral functions at fixed k-points in Fig.~\ref{fig:fig2}(c)-(f) and ~\ref{fig:fig3}(b)(c) 100 supercells of 800 atoms on average are used). 
The spectral functions are calculated directly from the supercell eigenvalues and eigenvectors, by applying the recently developed unfolding method~\cite{unfolding}.
The first principles simulations of the large supercells, necessary for a proper treatment of disorder, are made affordable by the recently developed Wannier function~\cite{wannier} based effective Hamiltonian method for disordered systems~\cite{naxco2}.
The influence of the TM substitutions on the Hamiltonian is extracted from three DFT~\cite{sup, blaha, i4mmm} calculations: the undoped BaFe$_2$As$_2$ and the impurity supercells Ba$_2$Fe$_3$CoAs$_4$ and Ba$_8$Fe$_{15}$ZnAs$_{16}$.
The low energy Hilbert space is taken within [-10,3] eV consisting of Wannier orbitals of Fe-$d$, Co-$d$/Zn-$d$, and As-$p$ characters. 
The accuracy of the calculated effective Hamiltonian is benchmarked~\cite{sup} against full DFT calculations.

Let us first re-examine the previously reported charge localization~\cite{wadati}
 and chemical potential shift~\cite{nakamura, konbu} within our disordered Co substitution case.
Fig.~\ref{fig:fig1}(b) shows the average occupation of each atom calculated from frequency/momentum integration and orbital summation of $\langle A_{n}(k,\omega)\rangle$.
Comparing the occupations in the doped system with that in the undoped system shows that, in agreement with Ref.~\cite{wadati,nakamura, konbu}, the additional charge is indeed distributed at the Co impurity, seemingly implying that Co does not dope carriers.
On the other hand, compared with the FSs of the undoped (Fig.~\ref{fig:fig1}(c)) system, the doped system (Fig.~\ref{fig:fig1}(d)) contains smaller/larger hole/electron pockets around the $\Gamma$/X point, in agreement with the previous experimental and theoretical findings ~\cite{neupane,malaeb,nakamura,konbu,liu}.
Furthermore the FSs of the doped system are found to be consistent with the undoped system upon adding the nominal doping of +0.125 electrons per Fe (the dotted line in Fig.~\ref{fig:fig1}(d).)
The same can also be observed from the band structure in Fig.~\ref{fig:fig2}(b).
So, our results of disordered Co substitutions confirm both the localized distribution of additional charge~\cite{wadati,nakamura,konbu} and the chemical potential shift~\cite{nakamura,konbu}, which still leaves us with the question: Do TM substitutions dope carriers or not? 

To reconcile these seemingly contradicting results, a notion of impurity potential~\cite{nakamura,konbu} (or self-energy~\cite{Haverkort}) induced``effective doping'' was previously proposed~\cite{nakamura,konbu}.
However, our results give a different physical explanation and show that such notion is  not only unnecessary but also negligible in effect for the case of Co substitution.
Let's focus on the ($k$,$\omega$)-dependence of $\langle A_{n}(k,\omega)\rangle$ in Fig.~\ref{fig:fig2}.
Notice first that while the band dispersion resembles well the undoped system, important effects of disorder scattering can be observed.
For example, the Fe-$d$ bands (even at the chemical potential) develop line widths in both momentum and frequency, reflecting their finite mean free path and lifetime from scattering against Co impurities.
Also notice that besides these features, no apparent localized state (flat band) is found in Fig~\ref{fig:fig2} to host one electron per Co.
Therefore, the localized distribution of additional charge must originate from \textit{slight} enhancement of the wave function near the Co sites in \textit{a large number} of extended occupied states, instead of a small number of strongly localized or resonant states!
This deduction is further supported by the DOS in Fig.~\ref{fig:fig2}(a), in which the \textit{broad} Co spectral weight ($\times 7$ for better comparison) is shown to be transferred to slightly lower energy, in relation to the Fe spectral weight.
That is, contrary to the previous interpretations~\cite{wadati,nakamura,konbu} our result shows that practically no carriers are ``trapped'' around Co, but instead a large number of itinerant carriers conspire to enhance the charge distribution around Co in order to screen the additional proton.
Therefore, the notion of "effective doping" is unnecessary in this case, and in fact negligible in effect.
Since the change of the wave functions and the band dispersion are rather small in this simplest case, from any physically meaningful consideration concerning doping, Co substitution really just adds about one more itinerant electron per Co to the system.

However, these itinerant carriers are not all the same in nature, as some of these carriers lose their coherence due to scattering against disordered impurities.
The lack of a well-defined dispersion indicates diffusive propagation of these incoherent carriers, in contrast to the particle-like propagation of the coherent carriers.
A clear example of non-dispersive spectral weight can be found in band 12 around $k_0=\Gamma$. 
As illustrated in Fig.~\ref{fig:fig2}(c)(d), the band consists of 
two components: a peak (dashed line) that disperses from 0.6 to 0.7eV along $\Gamma$M and a non-dispersive satellite (solid line) that remains fixed a little above 0.8eV. 
Other incoherent features are more difficult to recognize due to the presence of multiple bands in the same frequency range. 
To resolve the different bands, in Fig.~\ref{fig:fig2}(c)-(f) and ~\ref{fig:fig3}(b)(c), we analyze the spectral function
$A_{j}(k,\omega)=-\frac{1}{\pi}{\rm Im} \langle kj|G(\omega)|kj\rangle$ on the basis of the undoped Bloch states $|kj\rangle$ of crystal momentum $k$ and band index $j$~\cite{sup}.
In Fig.~\ref{fig:fig2}(e)(f) an incoherent component in the hole pocket is illustrated. While along $\Gamma$M the main peak in band 11 disperses from 0.1 to -0.3eV, the satellite around -0.6eV remains fixed. Note also that while nearby in frequency, the non-dispersive satellite of band 11 is not part of the dispersing band 8. 
Due to the proximity of the incoherent features to the main peaks, an unambiguous quantification of the obviously reduced coherent spectral weight is challenging in this case.

From the perspective of disordered impurities, Co substitution of Fe introduces only a weak impurity potential $\sim$-0.5eV to the d-orbitals, much smaller than the bandwidth of the $d$-bands, and thus can be considered as a small perturbation to the system.
This is ultimately the reason why the Fermi surfaces resemble very much the undoped system added with nominal doping of +0.125 electrons per Fe (e.g. Fig.1(d)), even though Luttinger theorem should not apply to disordered systems.
Therefore, it is instructive to explore the Zn substitution of Fe, which represents the other limit with very strong impurity potential ($\sim$-7eV) much larger than the bandwidth of the $d$-bands. 

As expected, the strong impurity potential of Zn substitution is able to induce real localization, in great contrast to the Co substitution.
Fig.~\ref{fig:fig3}(a) shows our resulting band structure of Zn substituted system with flat core levels of Zn-$d$ character, which are good examples of strongly localized states created by the large impurity potential.
More interestingly, weakly localized As-states can also be observed clearly within [-6,-5] eV, right below the As-$p$ bands~\cite{aspband}.
The small region of $k$-space in which these weakly localized states reside reflects their large size compared to the 2-Fe unit cell (similar to the oxygen vacancy states in ZnO~\cite{herng}).

Due to the very strong impurity potential, the influence of the impurity can no longer be considered as a small perturbation, and consequently the counting of the carriers would become non-trivial.
Indeed, the broadening of the Fe-$d$ spectrum in Fig.~\ref{fig:fig3}(b) is significantly stronger than the previous case, indicating a much shorter lifetime and mean free path.
In fact, even As-$p$ bands scatter strongly in Fig.~\ref{fig:fig3}(a).
Now, from a comparison of the valence of Fe$^{2+}$ and Zn$^{2+}$, one could have expected that Zn substitutions do not dope carriers to the system.
This is obviously not the case, since the chemical potential has been shifted so much that the hole carriers are almost entirely removed from the system.
The new FSs in Fig.~\ref{fig:fig3}(d) appear to have been doped with $0.206$ electrons per Fe, a number that cannot be easily rationalized from simple counting of valence, in great contrast to the case of Co substitution.
This illustrates clearly the general inapplicability of Luttinger theorem to disordered systems.

Of course, the physically meaningful count should distinguish the coherent carriers from the incoherent features.
For this purpose we analyze the components of the Bloch based spectral function. 
Two examples are illustrated in Fig.~\ref{fig:fig3}(b)(c), representing the bands in the hole and electron pockets respectively. 
An integration of the  Bloch based spectral function within [-1,1]eV gives an estimated coherence factor of $\sim$0.85. 
This results in an effective doping of $\sim$+0.049 / $\sim$-0.126 per Fe in the electron/hole pockets (see Fig.~\ref{fig:fig3}(f)).
A simple Luttinger count would lead to a large overestimation of 73\% of the number of doped electron carriers.

Interestingly, the Co/Zn substitutions are more efficient in removing the coherent hole carriers than adding the electron ones, an effect relevant to the doping dependence of Hall coefficient~\cite{fang, rullier,li}.
Of the $\sim 0.15$ spectral weight loss with Zn substitutions, $\sim 0.09$ comes from the trivial strongly localized Zn d-orbitals, while 
the remaining $\sim0.06$ corresponds to the decoherence of Fe bands.
(Similar spectral weight losses can be expected in the case of Fe vacancies.)
Finally, Fig.~\ref{fig:fig3}(e) shows an enhanced DOS upon substitution, even though the coherent fraction of DOS actually decreases.
This is quite alarming, considering that DOS is routinely used in the standard studies of doped materials, but actually provides no distinguishability of coherent features from the incoherent ones.

The above disorder effects have important physical implications that immediately impact on experimental and theoretical studies of FeSCs.
At the simplest level, one expects a doping dependence of the broadening of the bands, as found in the raw ARPES data~\cite{malaeb,liu}, and contribution to the incoherent features in other spectroscopies~\cite{lucarelli,nakajima,barisic}.
But, deeper issues are present.
As demonstrated above, other than the appearance of incoherent features, the chemical potential and the associated phase space of coherent carriers are no longer related to the coherent carrier density.
Indeed, we have found the Fermi velocities of the electron/hole carriers to resemble those of a highly doped system, but instead with a much smaller coherent carrier density.
This cannot be possibly described with merely adding carriers to the undoped system, and is in contradiction with the textbook theories, based on which most experiments are analyzed.
For instance, in optical conductivity and dielectric function, one might now find a spectral line shape similar to that of an undoped system with additional electrons (since its band dispersion resembles the real disordered system), but with a overall reduction in the intensity (due to the weight loss of coherent carriers) as dictated by various sum rules.
The same consideration should apply to transport and other spectroscopies as well, since they are all sensitive to both the phase space controlled by the band structure and the spectral weight related to the coherent carrier density.

More importantly, besides the impurities' standard role of suppressing $T_{c}$ via scattering~\cite{abrikosov,golubov,efremov}, the emergence of incoherent carriers and the lost of coherent spectral weight can also affect the competition between magnetic and superconducting instabilities in the underdoped regime.
Both effects serve to suppress long-range antiferromagnetic order, allowing a stronger superconductivity to develop.
This is because the the nesting condition is reduced by the loss of coherent spectral weight, while the ill-defined momentum of the incoherent carriers give negligible contribution.
This explains, at least in part, why both electron doping and hole doping both suppress the antiferromagnetic order, despite the degree of nesting condition.
On the other hand, the zero momentum superconductivity still benefit from all the incoherent carriers.
In fact, the slow propagation of the incoherent carriers makes them more susceptible to the strong coupling regime of superconductivity~\cite{Yildirim} that hosts stronger local pairing but weaker phase coherence.
(This might offer a resolution to the current issue of power law behavior of the penetration depth~\cite{prozorov}).
Coupled to the lighter coherent carriers, this can actually help enhance the superconductivity by providing a stronger local pairing.
Such opposite effects on the magnetism and superconductivity have in fact been observed recently in Zn substitution of F-underdoped LaOFeAs~\cite{li}.
Finally, the loss of coherent spectral weight would allow practically the current itinerant pairing theories~\cite{chubukov,hirschfeld,kunes,tesanovic} to incorporate stronger interaction parameters without triggering a long-range magnetic order, and thereby to further enhance the superconducting instability.

In conclusion, by computing the configurational-averaged spectral function of disordered Co/Zn substituted BaFe$_2$As$_2$, we identify important physical effects of TM substitution that cannot be captured by merely adding carriers to the undoped systems.
While both strongly and weakly localized states are found via Zn substitution, no apparent localization is found via Co substitution, despite the localized distribution of additional charge.
On the one hand, the FSs and Fermi velocities are modified as a consequence of chemical potential shift, reflecting injection of doped carriers.
On the other hand, some incoherent carriers emerge, in association with a loss of the coherent spectral weight that reduces considerably the actual number of coherent carriers.
Our results illustrate clearly that generally in disordered systems, none of the FSs, DOS, or the charge distribution can be solely used for the counting of coherent carriers.
Physically, our reported features would produce unusual behaviors in various spectroscopies and other physical properties beyond the standard textbook descriptions, and thus needs to be taken into account in future experimental and theoretical studies.
In addition, they offer new possibilities to tip the balance from magnetism toward superconductivity, an important aspect that deserves to be considered in future studies of superconductivity in general.

Work funded by the U S Department of Energy, Office of Basic Energy Sciences DE-AC02-98CH10886 and DOE-CMCSN.

\begin{widetext}
{\Large\bf Supplementary information}

\section{Details of the density functional theory calculations}\label{sec:secdft}

\begin{figure}[htp]
\includegraphics[width=0.85\columnwidth]{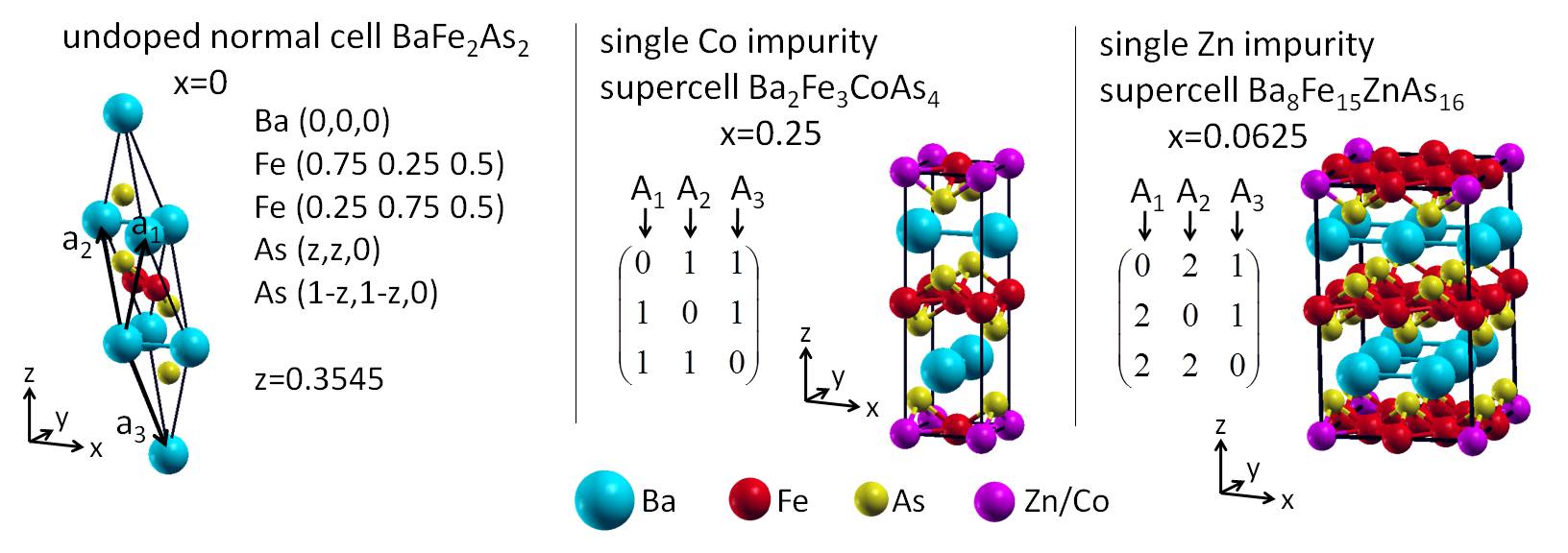}
\caption{\label{fig:figsup1}
(Left) undoped normal cell BaFe$_2$Co$_2$, (middle) single Co impurity supercell Ba$_2$Fe$_3$CoAs$_4$ and (right) single Zn impurity supercell Ba$_8$Fe$_{15}$ZnAs$_{16}$.}
\end{figure}

The lattice structure of the undoped normal cell, BaFe$_2$As$_2$, is depicted in the left of figure \ref{fig:figsup1}. Its space group I4/mmm and its lattice parameters a=7.49 bohr and c=24.60 bohr and z=0.3545, were taken from~\cite{i4mmm1}. The primitive normal cell lattice vectors expressed in Cartesian coordinates are given by:
$ a_1=-\frac{1}{2}a\hat{x}+\frac{1}{2}a\hat{y}+\frac{1}{2}c\hat{z} \; ; \;  a_2=\frac{1}{2}a\hat{x}-\frac{1}{2}a\hat{y}+\frac{1}{2}c\hat{z} \; ; \; a_3=\frac{1}{2}a\hat{x}+\frac{1}{2}a\hat{y}-\frac{1}{2}c\hat{z}$. 
The supercell used to capture the influence of the Co impurity Ba$_2$Fe$_3$CoAs$_4$ is depicted in the middle of figure \ref{fig:figsup1} and its primitive lattice vectors expressed in normal cell lattice vectors are given by:
$ A_1=a_2+a_3 \; ; \;  A_2=a_1+a_3  \; ; \; A_3=a_1+a_2 $. 
To capture the influence of the Zn impurity a four times larger supercell is used to properly treat its significant non-local influence on the nearest As and Fe orbitals. The Ba$_8$Fe$_{15}$ZnAs$_{16}$  supercell is depicted on the right of figure \ref{fig:figsup1} and its primitive lattice vectors expressed in normal cell lattice vectors are given by:
$ A_1=2a_2+2a_3 \; ; \;  A_2=2a_1+2a_3  \; ; \; A_3=a_1+a_2 $. 
We applied the WIEN2K\cite{blaha1} implementation of the full potential linearized augmented plane wave method in the
local density approximation. 
The k-point mesh was taken to be 10$\times$10$\times$10 for the undoped normal cell and 10$\times$10$\times$3 and 5$\times$5$\times$3 for the Co and Zn supercells respectively.
The basis set sizes were determined by RKmax=7. 

\section{Benchmarks of the effective Hamiltonian against DFT}

To explore the accuracy and efficiency of the effective Hamiltonian for the case of Ba(Fe$_{1-x}$M$_{x}$)$_{2}$As$_{2}$ with M=Co/Zn, we present comparisons of spectral functions $A_n(k,\omega)$  calculated from the full DFT and the effective Hamiltonian (see figures \ref{fig:figsup2a}-\ref{fig:figsup2c}). The size of the deviations between the full DFT and the effective Hamiltonian should be compared with the size of the impurity induced changes. For this purpose the spectral function of the undoped BaFe$_2$As$_2$ is also plotted as a reference for each benchmark. The test cases in figures \ref{fig:figsup2a} and \ref{fig:figsup2b} are designed as extreme test cases for the ``linearity'' approximation of the impurity influence (see formula (1) of Ref.~\cite{naxco21}). The case of BaCo$_2$As$_2$ corresponds to the maximum extrapolation, from $x=0.25$ in the single impurity cell (see middle of Fig. \ref{fig:figsup1}) to $x=1$. For the Zn doping, the BaFeZnAs$_2$ system is more meaningful case then BaZn$_2$As$_2$, in order to have remaining Fe bands. In this benchmark the dispersion of the Zn bands around -8 eV are less satisfactory captured by the effective Hamiltonian.  This can be understood as a good example in which additional two-impurity effects are needed in Eq.(1) of Ref.~\cite{naxco21}.  Nonetheless, the more relevant Fe bands near the chemical potential are very reasonably described, despite the extreme extrapolation in doping from $x=0.0625$ in the single impurity cell (see right of Fig. \ref{fig:figsup1}) to $x=0.5$. The test case in figure \ref{fig:figsup2c} of a $\sqrt{5}\times\sqrt{5}$ supercell is designed to test the partitioning of the impurity influence from its super images (see section IV of the supplementary of Ref.~\cite{naxco21}). The basis set of Linear Augmented Plane Waves (LAPW's) used in the full DFT is $\sim30$ times larger then the basis set of Wannier functions used in the effective Hamiltonian method. Since the number of floating point operations of diagonalization depends cubically on the size of the matrix this implies an efficiency increase by a factor of $30^{3}\approx3\cdot10^{4}$. Furthermore the full DFT calculation involves multiple self consistent cycles (typically 15) whereas the effective Hamiltonian method (as currently implemented) requires only a single diagonalization, which increases the efficiency by another order of magnitude to $\sim 6\cdot10^{5}$.

\begin{figure}[htp]
\includegraphics[width=0.8\columnwidth]{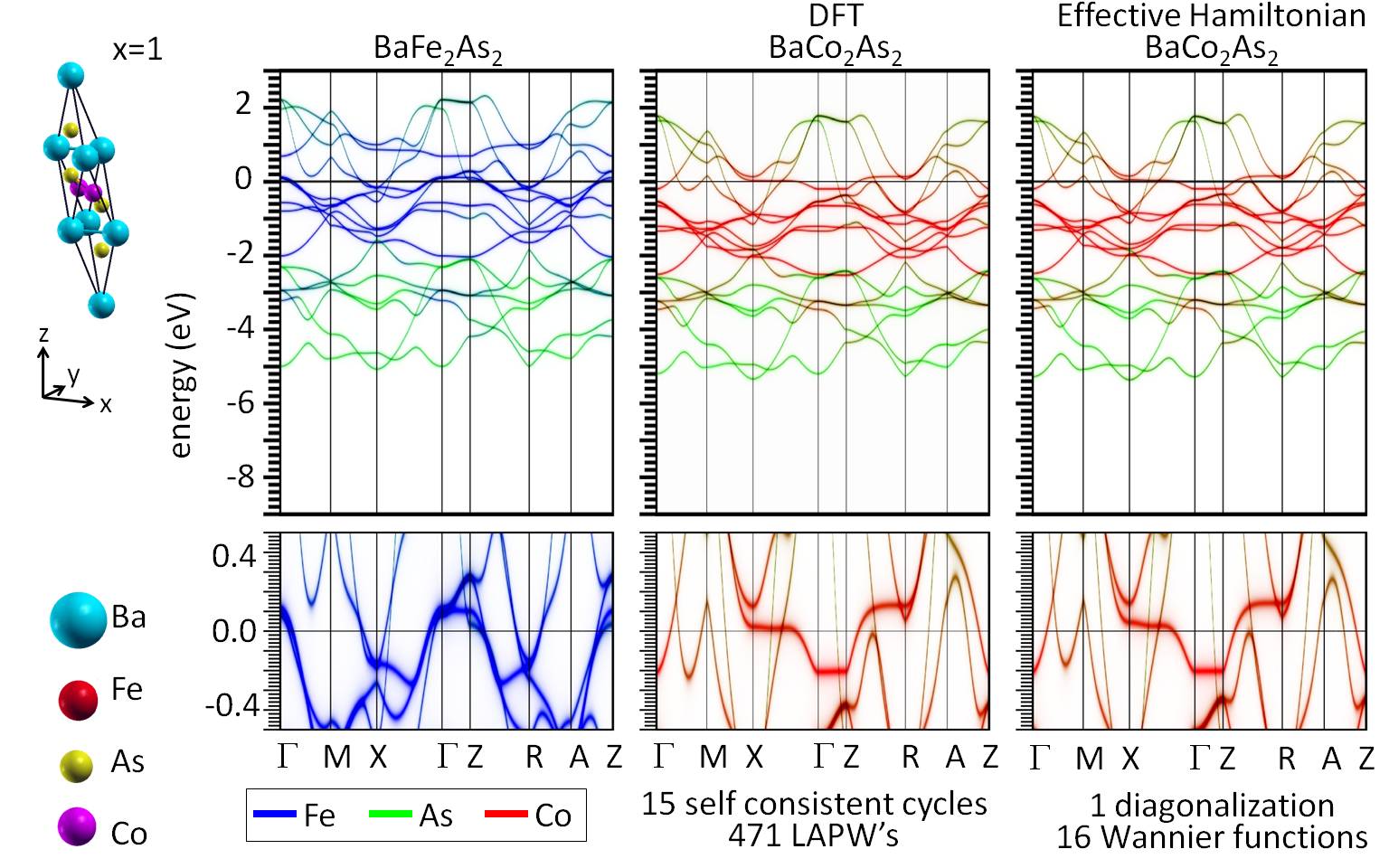}
\caption{\label{fig:figsup2a}
}
\end{figure}

\begin{figure}[htp]
\includegraphics[width=0.8\columnwidth]{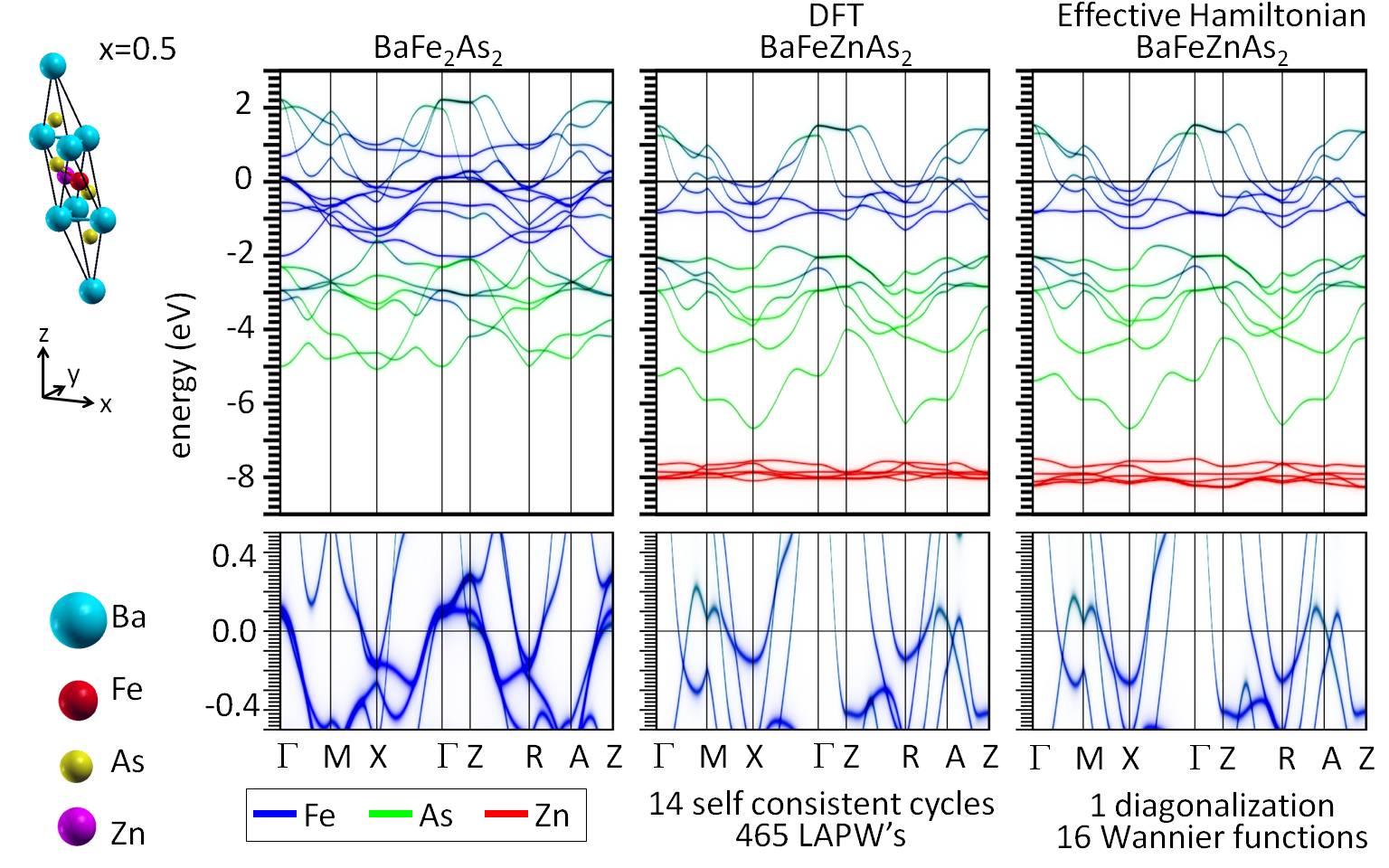}
\caption{\label{fig:figsup2b}
}
\end{figure}

\begin{figure}[htp]
\includegraphics[width=0.8\columnwidth]{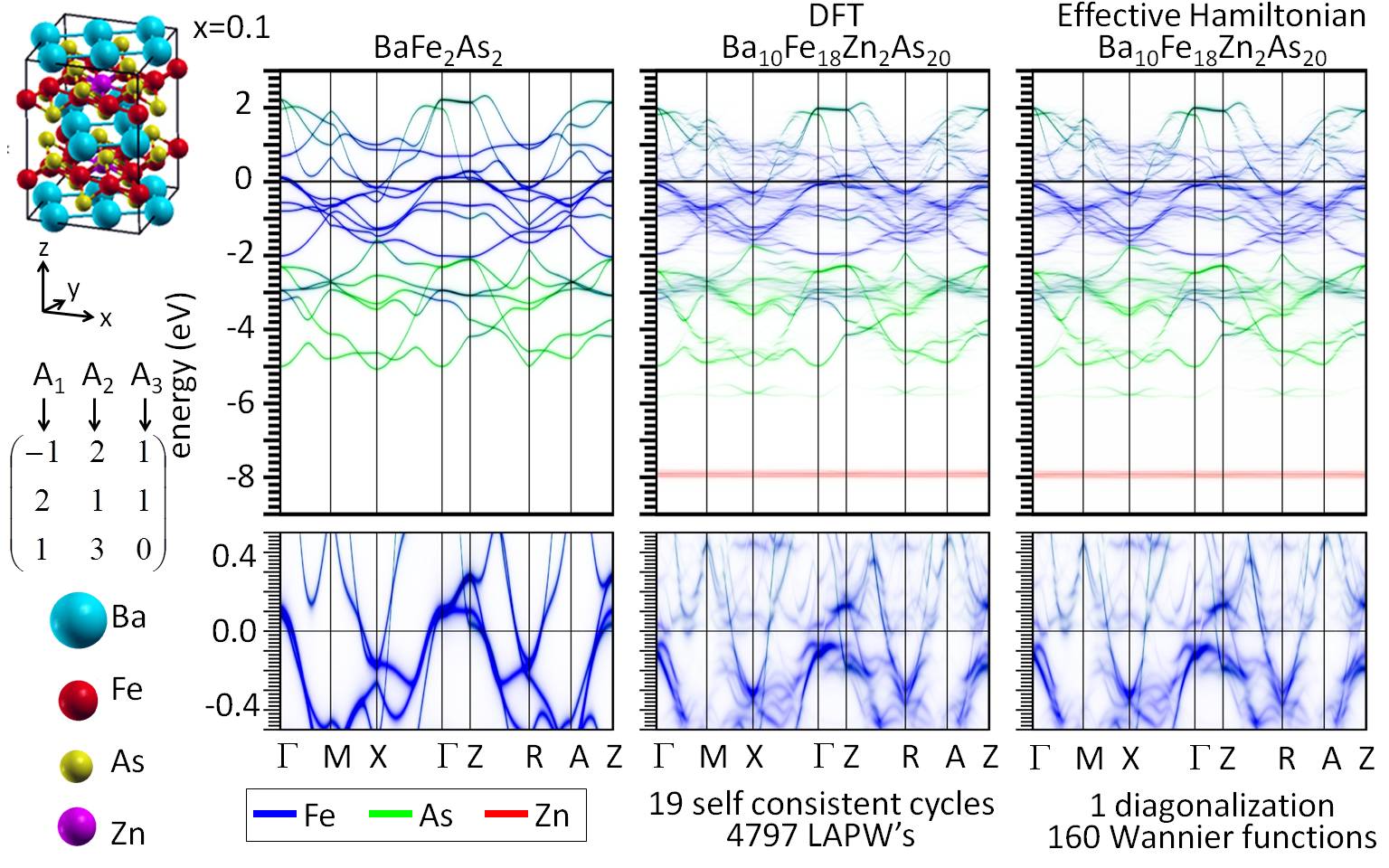}
\caption{\label{fig:figsup2c}
}
\end{figure}

\clearpage
\section{Convergence with respect to size and number of configurations}

In figures \ref{fig:figsup3a}/\ref{fig:figsup3b} we demonstrate the convergence of the configurational averaged spectral function $\langle A(k=k_0,\omega)\rangle$  of Ba(Fe$_{0.875}$Zn$_{0.125}$)$_{2}$As$_{2}$ at $k_0=\Gamma$/$k_0=\frac{2}{3}MX$, corresponding to the hole pocket/weakly localized As states.

\begin{figure}[htp]
\includegraphics[width=1\columnwidth]{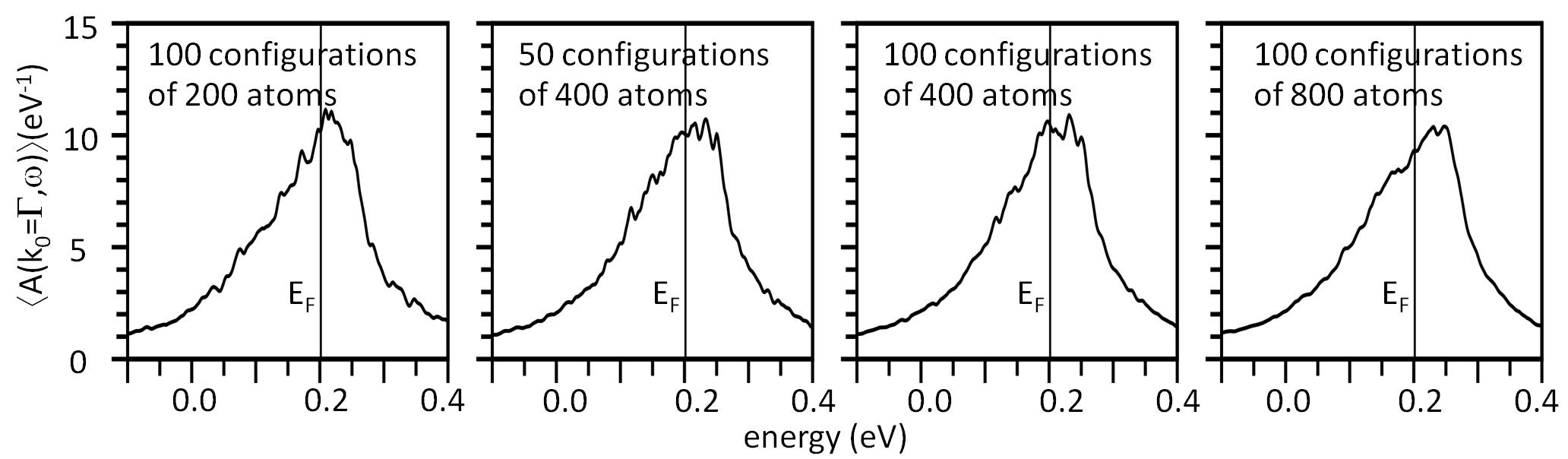}
\caption{\label{fig:figsup3a}
Demonstration of convergence of the configurational averaged spectral function of
Ba(Fe$_{0.875}$Zn$_{0.125}$)$_{2}$As$_{2}$ for the energy distribution curve $\langle A(k=k_0,\omega)\rangle$ at $k_0=\Gamma$. 
}
\end{figure}

\begin{figure}[htp]
\includegraphics[width=1\columnwidth]{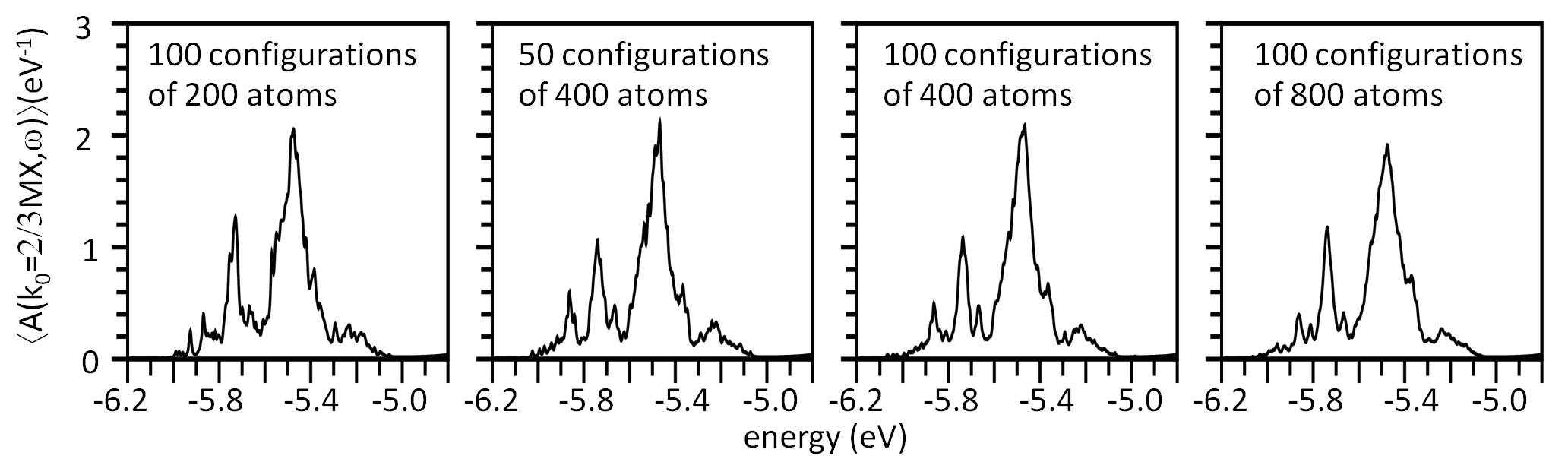}
\caption{\label{fig:figsup3b}
Demonstration of convergence of the configurational averaged spectral function of
Ba(Fe$_{0.875}$Zn$_{0.125}$)$_{2}$As$_{2}$ for the energy distribution curve $\langle A(k=k_0,\omega)\rangle$ at $k_0=\frac{2}{3}$MX. 
}
\end{figure}

\clearpage
\section{Bloch based spectral function}

The Bloch based spectral function is obtained from:
\begin{eqnarray}
A_{j}(k,\omega)=-\frac{1}{\pi}{\rm Im} \langle kj|G(\omega)|kj\rangle=\sum_{J}\delta(\omega-\epsilon_{KJ}) |\langle kj|KJ\rangle|^2\; ; \; 
\langle kj|KJ\rangle=\sum_{n}\langle kj|kn\rangle \langle kn |KJ\rangle
\end{eqnarray}
with $k/K$ is the normal/super cell crystal momentum, $\omega$ the frequency, $j/J$ the normal/super cell band index, $n$ the normal cell Wannier index. Here $\epsilon_{KJ}$ and $|KJ\rangle$ are the supercell eigenvalue and eigenstate, $\langle kn|kj\rangle$ is the Wannier based Bloch state of the undoped system and $\langle kn |KJ\rangle$ is the unfolding matrix element (see formula (4) of Ref.~\cite{unfolding1}). In figure \ref{fig:figsup6} we display the Bloch based spectral function for Ba(Fe$_{0.875}$Zn$_{0.125}$)$_{2}$As$_{2}$ at $k_0=\frac{2}{3}MX$.

\begin{figure}[htp]
\includegraphics[width=1\columnwidth]{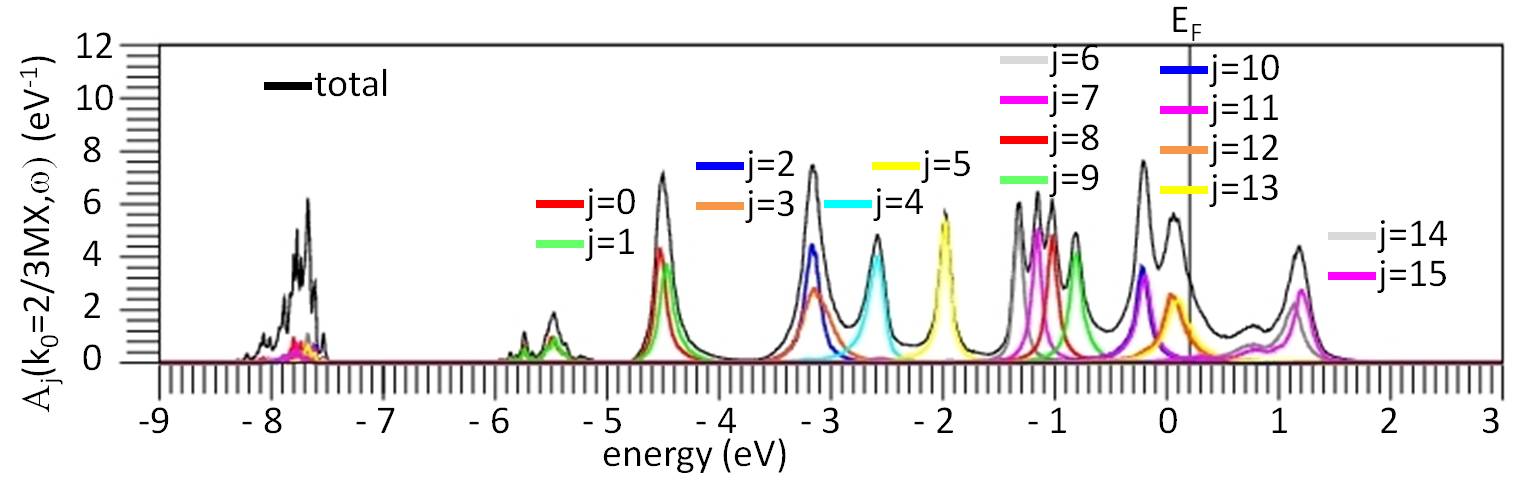}
\caption{\label{fig:figsup6}
Bloch based spectral function $\langle A_j(k=k_0,\omega)\rangle$ of band index $j$ for Ba(Fe$_{0.875}$Zn$_{0.125}$)$_{2}$As$_{2}$ at the fixed k-point $k_0=\frac{2}{3}MX$
}
\end{figure}

\end{widetext}


\begin{thebibliography}{00}
\bibitem{aspband} This does not rule out potential diffusive itinerancy near the center of the impurity band.
\end{thebibliography}
\end{document}